\documentclass[useAMS,usegraphicx,usenatbib]{mn2e} 

\usepackage{ulem}
\usepackage{color}
\definecolor{AliceBlue}{rgb}{0.94,0.97,1.00}
\definecolor{AntiqueWhite1}{rgb}{1.00,0.94,0.86}
\definecolor{AntiqueWhite2}{rgb}{0.93,0.87,0.80}
\definecolor{AntiqueWhite3}{rgb}{0.80,0.75,0.69}
\definecolor{AntiqueWhite4}{rgb}{0.55,0.51,0.47}
\definecolor{AntiqueWhite}{rgb}{0.98,0.92,0.84}
\definecolor{BlanchedAlmond}{rgb}{1.00,0.92,0.80}
\definecolor{BlueViolet}{rgb}{0.54,0.17,0.89}
\definecolor{CadetBlue1}{rgb}{0.60,0.96,1.00}
\definecolor{CadetBlue2}{rgb}{0.56,0.90,0.93}
\definecolor{CadetBlue3}{rgb}{0.48,0.77,0.80}
\definecolor{CadetBlue4}{rgb}{0.33,0.53,0.55}
\definecolor{CadetBlue}{rgb}{0.37,0.62,0.63}
\definecolor{CornflowerBlue}{rgb}{0.39,0.58,0.93}
\definecolor{DarkBlue}{rgb}{0.00,0.00,0.55}
\definecolor{DarkCyan}{rgb}{0.00,0.55,0.55}
\definecolor{DarkGoldenrod1}{rgb}{1.00,0.73,0.06}
\definecolor{DarkGoldenrod2}{rgb}{0.93,0.68,0.05}
\definecolor{DarkGoldenrod3}{rgb}{0.80,0.58,0.05}
\definecolor{DarkGoldenrod4}{rgb}{0.55,0.40,0.03}
\definecolor{DarkGoldenrod}{rgb}{0.72,0.53,0.04}
\definecolor{DarkGray}{rgb}{0.66,0.66,0.66}
\definecolor{DarkGreen}{rgb}{0.00,0.39,0.00}
\definecolor{DarkGrey}{rgb}{0.66,0.66,0.66}
\definecolor{DarkKhaki}{rgb}{0.74,0.72,0.42}
\definecolor{DarkMagenta}{rgb}{0.55,0.00,0.55}
\definecolor{DarkOliveGreen1}{rgb}{0.79,1.00,0.44}
\definecolor{DarkOliveGreen2}{rgb}{0.74,0.93,0.41}
\definecolor{DarkOliveGreen3}{rgb}{0.64,0.80,0.35}
\definecolor{DarkOliveGreen4}{rgb}{0.43,0.55,0.24}
\definecolor{DarkOliveGreen}{rgb}{0.33,0.42,0.18}
\definecolor{DarkOrange1}{rgb}{1.00,0.50,0.00}
\definecolor{DarkOrange2}{rgb}{0.93,0.46,0.00}
\definecolor{DarkOrange3}{rgb}{0.80,0.40,0.00}
\definecolor{DarkOrange4}{rgb}{0.55,0.27,0.00}
\definecolor{DarkOrange}{rgb}{1.00,0.55,0.00}
\definecolor{DarkOrchid1}{rgb}{0.75,0.24,1.00}
\definecolor{DarkOrchid2}{rgb}{0.70,0.23,0.93}
\definecolor{DarkOrchid3}{rgb}{0.60,0.20,0.80}
\definecolor{DarkOrchid4}{rgb}{0.41,0.13,0.55}
\definecolor{DarkOrchid}{rgb}{0.60,0.20,0.80}
\definecolor{DarkRed}{rgb}{0.55,0.00,0.00}
\definecolor{DarkSalmon}{rgb}{0.91,0.59,0.48}
\definecolor{DarkSeaGreen1}{rgb}{0.76,1.00,0.76}
\definecolor{DarkSeaGreen2}{rgb}{0.71,0.93,0.71}
\definecolor{DarkSeaGreen3}{rgb}{0.61,0.80,0.61}
\definecolor{DarkSeaGreen4}{rgb}{0.41,0.55,0.41}
\definecolor{DarkSeaGreen}{rgb}{0.56,0.74,0.56}
\definecolor{DarkSlateBlue}{rgb}{0.28,0.24,0.55}
\definecolor{DarkSlateGray1}{rgb}{0.59,1.00,1.00}
\definecolor{DarkSlateGray2}{rgb}{0.55,0.93,0.93}
\definecolor{DarkSlateGray3}{rgb}{0.47,0.80,0.80}
\definecolor{DarkSlateGray4}{rgb}{0.32,0.55,0.55}
\definecolor{DarkSlateGray}{rgb}{0.18,0.31,0.31}
\definecolor{DarkSlateGrey}{rgb}{0.18,0.31,0.31}
\definecolor{DarkTurquoise}{rgb}{0.00,0.81,0.82}
\definecolor{DarkViolet}{rgb}{0.58,0.00,0.83}
\definecolor{DeepPink1}{rgb}{1.00,0.08,0.58}
\definecolor{DeepPink2}{rgb}{0.93,0.07,0.54}
\definecolor{DeepPink3}{rgb}{0.80,0.06,0.46}
\definecolor{DeepPink4}{rgb}{0.55,0.04,0.31}
\definecolor{DeepPink}{rgb}{1.00,0.08,0.58}
\definecolor{DeepSkyBlue1}{rgb}{0.00,0.75,1.00}
\definecolor{DeepSkyBlue2}{rgb}{0.00,0.70,0.93}
\definecolor{DeepSkyBlue3}{rgb}{0.00,0.60,0.80}
\definecolor{DeepSkyBlue4}{rgb}{0.00,0.41,0.55}
\definecolor{DeepSkyBlue}{rgb}{0.00,0.75,1.00}
\definecolor{DimGray}{rgb}{0.41,0.41,0.41}
\definecolor{DimGrey}{rgb}{0.41,0.41,0.41}
\definecolor{DodgerBlue1}{rgb}{0.12,0.56,1.00}
\definecolor{DodgerBlue2}{rgb}{0.11,0.53,0.93}
\definecolor{DodgerBlue3}{rgb}{0.09,0.45,0.80}
\definecolor{DodgerBlue4}{rgb}{0.06,0.31,0.55}
\definecolor{DodgerBlue}{rgb}{0.12,0.56,1.00}
\definecolor{FloralWhite}{rgb}{1.00,0.98,0.94}
\definecolor{ForestGreen}{rgb}{0.13,0.55,0.13}
\definecolor{GhostWhite}{rgb}{0.97,0.97,1.00}
\definecolor{GreenYellow}{rgb}{0.68,1.00,0.18}
\definecolor{HotPink1}{rgb}{1.00,0.43,0.71}
\definecolor{HotPink2}{rgb}{0.93,0.42,0.65}
\definecolor{HotPink3}{rgb}{0.80,0.38,0.56}
\definecolor{HotPink4}{rgb}{0.55,0.23,0.38}
\definecolor{HotPink}{rgb}{1.00,0.41,0.71}
\definecolor{IndianRed1}{rgb}{1.00,0.42,0.42}
\definecolor{IndianRed2}{rgb}{0.93,0.39,0.39}
\definecolor{IndianRed3}{rgb}{0.80,0.33,0.33}
\definecolor{IndianRed4}{rgb}{0.55,0.23,0.23}
\definecolor{IndianRed}{rgb}{0.80,0.36,0.36}
\definecolor{LavenderBlush1}{rgb}{1.00,0.94,0.96}
\definecolor{LavenderBlush2}{rgb}{0.93,0.88,0.90}
\definecolor{LavenderBlush3}{rgb}{0.80,0.76,0.77}
\definecolor{LavenderBlush4}{rgb}{0.55,0.51,0.53}
\definecolor{LavenderBlush}{rgb}{1.00,0.94,0.96}
\definecolor{LawnGreen}{rgb}{0.49,0.99,0.00}
\definecolor{LemonChiffon1}{rgb}{1.00,0.98,0.80}
\definecolor{LemonChiffon2}{rgb}{0.93,0.91,0.75}
\definecolor{LemonChiffon3}{rgb}{0.80,0.79,0.65}
\definecolor{LemonChiffon4}{rgb}{0.55,0.54,0.44}
\definecolor{LemonChiffon}{rgb}{1.00,0.98,0.80}
\definecolor{LightBlue1}{rgb}{0.75,0.94,1.00}
\definecolor{LightBlue2}{rgb}{0.70,0.87,0.93}
\definecolor{LightBlue3}{rgb}{0.60,0.75,0.80}
\definecolor{LightBlue4}{rgb}{0.41,0.51,0.55}
\definecolor{LightBlue}{rgb}{0.68,0.85,0.90}
\definecolor{LightCoral}{rgb}{0.94,0.50,0.50}
\definecolor{LightCyan1}{rgb}{0.88,1.00,1.00}
\definecolor{LightCyan2}{rgb}{0.82,0.93,0.93}
\definecolor{LightCyan3}{rgb}{0.71,0.80,0.80}
\definecolor{LightCyan4}{rgb}{0.48,0.55,0.55}
\definecolor{LightCyan}{rgb}{0.88,1.00,1.00}
\definecolor{LightGoldenrod1}{rgb}{1.00,0.93,0.55}
\definecolor{LightGoldenrod2}{rgb}{0.93,0.86,0.51}
\definecolor{LightGoldenrod3}{rgb}{0.80,0.75,0.44}
\definecolor{LightGoldenrod4}{rgb}{0.55,0.51,0.30}
\definecolor{LightGoldenrodYellow}{rgb}{0.98,0.98,0.82}
\definecolor{LightGoldenrod}{rgb}{0.93,0.87,0.51}
\definecolor{LightGray}{rgb}{0.83,0.83,0.83}
\definecolor{LightGreen}{rgb}{0.56,0.93,0.56}
\definecolor{LightGrey}{rgb}{0.83,0.83,0.83}
\definecolor{LightPink1}{rgb}{1.00,0.68,0.73}
\definecolor{LightPink2}{rgb}{0.93,0.64,0.68}
\definecolor{LightPink3}{rgb}{0.80,0.55,0.58}
\definecolor{LightPink4}{rgb}{0.55,0.37,0.40}
\definecolor{LightPink}{rgb}{1.00,0.71,0.76}
\definecolor{LightSalmon1}{rgb}{1.00,0.63,0.48}
\definecolor{LightSalmon2}{rgb}{0.93,0.58,0.45}
\definecolor{LightSalmon3}{rgb}{0.80,0.51,0.38}
\definecolor{LightSalmon4}{rgb}{0.55,0.34,0.26}
\definecolor{LightSalmon}{rgb}{1.00,0.63,0.48}
\definecolor{LightSeaGreen}{rgb}{0.13,0.70,0.67}
\definecolor{LightSkyBlue1}{rgb}{0.69,0.89,1.00}
\definecolor{LightSkyBlue2}{rgb}{0.64,0.83,0.93}
\definecolor{LightSkyBlue3}{rgb}{0.55,0.71,0.80}
\definecolor{LightSkyBlue4}{rgb}{0.38,0.48,0.55}
\definecolor{LightSkyBlue}{rgb}{0.53,0.81,0.98}
\definecolor{LightSlateBlue}{rgb}{0.52,0.44,1.00}
\definecolor{LightSlateGray}{rgb}{0.47,0.53,0.60}
\definecolor{LightSlateGrey}{rgb}{0.47,0.53,0.60}
\definecolor{LightSteelBlue1}{rgb}{0.79,0.88,1.00}
\definecolor{LightSteelBlue2}{rgb}{0.74,0.82,0.93}
\definecolor{LightSteelBlue3}{rgb}{0.64,0.71,0.80}
\definecolor{LightSteelBlue4}{rgb}{0.43,0.48,0.55}
\definecolor{LightSteelBlue}{rgb}{0.69,0.77,0.87}
\definecolor{LightYellow1}{rgb}{1.00,1.00,0.88}
\definecolor{LightYellow2}{rgb}{0.93,0.93,0.82}
\definecolor{LightYellow3}{rgb}{0.80,0.80,0.71}
\definecolor{LightYellow4}{rgb}{0.55,0.55,0.48}
\definecolor{LightYellow}{rgb}{1.00,1.00,0.88}
\definecolor{LimeGreen}{rgb}{0.20,0.80,0.20}
\definecolor{MediumAquamarine}{rgb}{0.40,0.80,0.67}
\definecolor{MediumBlue}{rgb}{0.00,0.00,0.80}
\definecolor{MediumOrchid1}{rgb}{0.88,0.40,1.00}
\definecolor{MediumOrchid2}{rgb}{0.82,0.37,0.93}
\definecolor{MediumOrchid3}{rgb}{0.71,0.32,0.80}
\definecolor{MediumOrchid4}{rgb}{0.48,0.22,0.55}
\definecolor{MediumOrchid}{rgb}{0.73,0.33,0.83}
\definecolor{MediumPurple1}{rgb}{0.67,0.51,1.00}
\definecolor{MediumPurple2}{rgb}{0.62,0.47,0.93}
\definecolor{MediumPurple3}{rgb}{0.54,0.41,0.80}
\definecolor{MediumPurple4}{rgb}{0.36,0.28,0.55}
\definecolor{MediumPurple}{rgb}{0.58,0.44,0.86}
\definecolor{MediumSeaGreen}{rgb}{0.24,0.70,0.44}
\definecolor{MediumSlateBlue}{rgb}{0.48,0.41,0.93}
\definecolor{MediumSpringGreen}{rgb}{0.00,0.98,0.60}
\definecolor{MediumTurquoise}{rgb}{0.28,0.82,0.80}
\definecolor{MediumVioletRed}{rgb}{0.78,0.08,0.52}
\definecolor{MidnightBlue}{rgb}{0.10,0.10,0.44}
\definecolor{MintCream}{rgb}{0.96,1.00,0.98}
\definecolor{MistyRose1}{rgb}{1.00,0.89,0.88}
\definecolor{MistyRose2}{rgb}{0.93,0.84,0.82}
\definecolor{MistyRose3}{rgb}{0.80,0.72,0.71}
\definecolor{MistyRose4}{rgb}{0.55,0.49,0.48}
\definecolor{MistyRose}{rgb}{1.00,0.89,0.88}
\definecolor{NavajoWhite1}{rgb}{1.00,0.87,0.68}
\definecolor{NavajoWhite2}{rgb}{0.93,0.81,0.63}
\definecolor{NavajoWhite3}{rgb}{0.80,0.70,0.55}
\definecolor{NavajoWhite4}{rgb}{0.55,0.47,0.37}
\definecolor{NavajoWhite}{rgb}{1.00,0.87,0.68}
\definecolor{NavyBlue}{rgb}{0.00,0.00,0.50}
\definecolor{OldLace}{rgb}{0.99,0.96,0.90}
\definecolor{OliveDrab1}{rgb}{0.75,1.00,0.24}
\definecolor{OliveDrab2}{rgb}{0.70,0.93,0.23}
\definecolor{OliveDrab3}{rgb}{0.60,0.80,0.20}
\definecolor{OliveDrab4}{rgb}{0.41,0.55,0.13}
\definecolor{OliveDrab}{rgb}{0.42,0.56,0.14}
\definecolor{OrangeRed1}{rgb}{1.00,0.27,0.00}
\definecolor{OrangeRed2}{rgb}{0.93,0.25,0.00}
\definecolor{OrangeRed3}{rgb}{0.80,0.22,0.00}
\definecolor{OrangeRed4}{rgb}{0.55,0.15,0.00}
\definecolor{OrangeRed}{rgb}{1.00,0.27,0.00}
\definecolor{PaleGoldenrod}{rgb}{0.93,0.91,0.67}
\definecolor{PaleGreen1}{rgb}{0.60,1.00,0.60}
\definecolor{PaleGreen2}{rgb}{0.56,0.93,0.56}
\definecolor{PaleGreen3}{rgb}{0.49,0.80,0.49}
\definecolor{PaleGreen4}{rgb}{0.33,0.55,0.33}
\definecolor{PaleGreen}{rgb}{0.60,0.98,0.60}
\definecolor{PaleTurquoise1}{rgb}{0.73,1.00,1.00}
\definecolor{PaleTurquoise2}{rgb}{0.68,0.93,0.93}
\definecolor{PaleTurquoise3}{rgb}{0.59,0.80,0.80}
\definecolor{PaleTurquoise4}{rgb}{0.40,0.55,0.55}
\definecolor{PaleTurquoise}{rgb}{0.69,0.93,0.93}
\definecolor{PaleVioletRed1}{rgb}{1.00,0.51,0.67}
\definecolor{PaleVioletRed2}{rgb}{0.93,0.47,0.62}
\definecolor{PaleVioletRed3}{rgb}{0.80,0.41,0.54}
\definecolor{PaleVioletRed4}{rgb}{0.55,0.28,0.36}
\definecolor{PaleVioletRed}{rgb}{0.86,0.44,0.58}
\definecolor{PapayaWhip}{rgb}{1.00,0.94,0.84}
\definecolor{PeachPuff1}{rgb}{1.00,0.85,0.73}
\definecolor{PeachPuff2}{rgb}{0.93,0.80,0.68}
\definecolor{PeachPuff3}{rgb}{0.80,0.69,0.58}
\definecolor{PeachPuff4}{rgb}{0.55,0.47,0.40}
\definecolor{PeachPuff}{rgb}{1.00,0.85,0.73}
\definecolor{PowderBlue}{rgb}{0.69,0.88,0.90}
\definecolor{RosyBrown1}{rgb}{1.00,0.76,0.76}
\definecolor{RosyBrown2}{rgb}{0.93,0.71,0.71}
\definecolor{RosyBrown3}{rgb}{0.80,0.61,0.61}
\definecolor{RosyBrown4}{rgb}{0.55,0.41,0.41}
\definecolor{RosyBrown}{rgb}{0.74,0.56,0.56}
\definecolor{RoyalBlue1}{rgb}{0.28,0.46,1.00}
\definecolor{RoyalBlue2}{rgb}{0.26,0.43,0.93}
\definecolor{RoyalBlue3}{rgb}{0.23,0.37,0.80}
\definecolor{RoyalBlue4}{rgb}{0.15,0.25,0.55}
\definecolor{RoyalBlue}{rgb}{0.25,0.41,0.88}
\definecolor{SaddleBrown}{rgb}{0.55,0.27,0.07}
\definecolor{SandyBrown}{rgb}{0.96,0.64,0.38}
\definecolor{SeaGreen1}{rgb}{0.33,1.00,0.62}
\definecolor{SeaGreen2}{rgb}{0.31,0.93,0.58}
\definecolor{SeaGreen3}{rgb}{0.26,0.80,0.50}
\definecolor{SeaGreen4}{rgb}{0.18,0.55,0.34}
\definecolor{SeaGreen}{rgb}{0.18,0.55,0.34}
\definecolor{SkyBlue1}{rgb}{0.53,0.81,1.00}
\definecolor{SkyBlue2}{rgb}{0.49,0.75,0.93}
\definecolor{SkyBlue3}{rgb}{0.42,0.65,0.80}
\definecolor{SkyBlue4}{rgb}{0.29,0.44,0.55}
\definecolor{SkyBlue}{rgb}{0.53,0.81,0.92}
\definecolor{SlateBlue1}{rgb}{0.51,0.44,1.00}
\definecolor{SlateBlue2}{rgb}{0.48,0.40,0.93}
\definecolor{SlateBlue3}{rgb}{0.41,0.35,0.80}
\definecolor{SlateBlue4}{rgb}{0.28,0.24,0.55}
\definecolor{SlateBlue}{rgb}{0.42,0.35,0.80}
\definecolor{SlateGray1}{rgb}{0.78,0.89,1.00}
\definecolor{SlateGray2}{rgb}{0.73,0.83,0.93}
\definecolor{SlateGray3}{rgb}{0.62,0.71,0.80}
\definecolor{SlateGray4}{rgb}{0.42,0.48,0.55}
\definecolor{SlateGray}{rgb}{0.44,0.50,0.56}
\definecolor{SlateGrey}{rgb}{0.44,0.50,0.56}
\definecolor{SpringGreen1}{rgb}{0.00,1.00,0.50}
\definecolor{SpringGreen2}{rgb}{0.00,0.93,0.46}
\definecolor{SpringGreen3}{rgb}{0.00,0.80,0.40}
\definecolor{SpringGreen4}{rgb}{0.00,0.55,0.27}
\definecolor{SpringGreen}{rgb}{0.00,1.00,0.50}
\definecolor{SteelBlue1}{rgb}{0.39,0.72,1.00}
\definecolor{SteelBlue2}{rgb}{0.36,0.67,0.93}
\definecolor{SteelBlue3}{rgb}{0.31,0.58,0.80}
\definecolor{SteelBlue4}{rgb}{0.21,0.39,0.55}
\definecolor{SteelBlue}{rgb}{0.27,0.51,0.71}
\definecolor{VioletRed1}{rgb}{1.00,0.24,0.59}
\definecolor{VioletRed2}{rgb}{0.93,0.23,0.55}
\definecolor{VioletRed3}{rgb}{0.80,0.20,0.47}
\definecolor{VioletRed4}{rgb}{0.55,0.13,0.32}
\definecolor{VioletRed}{rgb}{0.82,0.13,0.56}
\definecolor{WhiteSmoke}{rgb}{0.96,0.96,0.96}
\definecolor{YellowGreen}{rgb}{0.60,0.80,0.20}
\definecolor{aliceblue}{rgb}{0.94,0.97,1.00}
\definecolor{antiquewhite}{rgb}{0.98,0.92,0.84}
\definecolor{aquamarine1}{rgb}{0.50,1.00,0.83}
\definecolor{aquamarine2}{rgb}{0.46,0.93,0.78}
\definecolor{aquamarine3}{rgb}{0.40,0.80,0.67}
\definecolor{aquamarine4}{rgb}{0.27,0.55,0.45}
\definecolor{aquamarine}{rgb}{0.50,1.00,0.83}
\definecolor{azure1}{rgb}{0.94,1.00,1.00}
\definecolor{azure2}{rgb}{0.88,0.93,0.93}
\definecolor{azure3}{rgb}{0.76,0.80,0.80}
\definecolor{azure4}{rgb}{0.51,0.55,0.55}
\definecolor{azure}{rgb}{0.94,1.00,1.00}
\definecolor{beige}{rgb}{0.96,0.96,0.86}
\definecolor{bisque1}{rgb}{1.00,0.89,0.77}
\definecolor{bisque2}{rgb}{0.93,0.84,0.72}
\definecolor{bisque3}{rgb}{0.80,0.72,0.62}
\definecolor{bisque4}{rgb}{0.55,0.49,0.42}
\definecolor{bisque}{rgb}{1.00,0.89,0.77}
\definecolor{black}{rgb}{0.00,0.00,0.00}
\definecolor{blanchedalmond}{rgb}{1.00,0.92,0.80}
\definecolor{blue1}{rgb}{0.00,0.00,1.00}
\definecolor{blue2}{rgb}{0.00,0.00,0.93}
\definecolor{blue3}{rgb}{0.00,0.00,0.80}
\definecolor{blue4}{rgb}{0.00,0.00,0.55}
\definecolor{blueviolet}{rgb}{0.54,0.17,0.89}
\definecolor{blue}{rgb}{0.00,0.00,1.00}
\definecolor{brown1}{rgb}{1.00,0.25,0.25}
\definecolor{brown2}{rgb}{0.93,0.23,0.23}
\definecolor{brown3}{rgb}{0.80,0.20,0.20}
\definecolor{brown4}{rgb}{0.55,0.14,0.14}
\definecolor{brown}{rgb}{0.65,0.16,0.16}
\definecolor{burlywood1}{rgb}{1.00,0.83,0.61}
\definecolor{burlywood2}{rgb}{0.93,0.77,0.57}
\definecolor{burlywood3}{rgb}{0.80,0.67,0.49}
\definecolor{burlywood4}{rgb}{0.55,0.45,0.33}
\definecolor{burlywood}{rgb}{0.87,0.72,0.53}
\definecolor{cadetblue}{rgb}{0.37,0.62,0.63}
\definecolor{chartreuse1}{rgb}{0.50,1.00,0.00}
\definecolor{chartreuse2}{rgb}{0.46,0.93,0.00}
\definecolor{chartreuse3}{rgb}{0.40,0.80,0.00}
\definecolor{chartreuse4}{rgb}{0.27,0.55,0.00}
\definecolor{chartreuse}{rgb}{0.50,1.00,0.00}
\definecolor{chocolate1}{rgb}{1.00,0.50,0.14}
\definecolor{chocolate2}{rgb}{0.93,0.46,0.13}
\definecolor{chocolate3}{rgb}{0.80,0.40,0.11}
\definecolor{chocolate4}{rgb}{0.55,0.27,0.07}
\definecolor{chocolate}{rgb}{0.82,0.41,0.12}
\definecolor{coral1}{rgb}{1.00,0.45,0.34}
\definecolor{coral2}{rgb}{0.93,0.42,0.31}
\definecolor{coral3}{rgb}{0.80,0.36,0.27}
\definecolor{coral4}{rgb}{0.55,0.24,0.18}
\definecolor{coral}{rgb}{1.00,0.50,0.31}
\definecolor{cornflowerblue}{rgb}{0.39,0.58,0.93}
\definecolor{cornsilk1}{rgb}{1.00,0.97,0.86}
\definecolor{cornsilk2}{rgb}{0.93,0.91,0.80}
\definecolor{cornsilk3}{rgb}{0.80,0.78,0.69}
\definecolor{cornsilk4}{rgb}{0.55,0.53,0.47}
\definecolor{cornsilk}{rgb}{1.00,0.97,0.86}
\definecolor{cyan1}{rgb}{0.00,1.00,1.00}
\definecolor{cyan2}{rgb}{0.00,0.93,0.93}
\definecolor{cyan3}{rgb}{0.00,0.80,0.80}
\definecolor{cyan4}{rgb}{0.00,0.55,0.55}
\definecolor{cyan}{rgb}{0.00,1.00,1.00}
\definecolor{darkblue}{rgb}{0.00,0.00,0.55}
\definecolor{darkcyan}{rgb}{0.00,0.55,0.55}
\definecolor{darkgoldenrod}{rgb}{0.72,0.53,0.04}
\definecolor{darkgray}{rgb}{0.66,0.66,0.66}
\definecolor{darkgreen}{rgb}{0.00,0.39,0.00}
\definecolor{darkgrey}{rgb}{0.66,0.66,0.66}
\definecolor{darkkhaki}{rgb}{0.74,0.72,0.42}
\definecolor{darkmagenta}{rgb}{0.55,0.00,0.55}
\definecolor{darkolive}{rgb}{0.33,0.42,0.18}
\definecolor{darkorange}{rgb}{1.00,0.55,0.00}
\definecolor{darkorchid}{rgb}{0.60,0.20,0.80}
\definecolor{darkred}{rgb}{0.55,0.00,0.00}
\definecolor{darksalmon}{rgb}{0.91,0.59,0.48}
\definecolor{darksea}{rgb}{0.56,0.74,0.56}
\definecolor{darkslate}{rgb}{0.18,0.31,0.31}
\definecolor{darkslate}{rgb}{0.18,0.31,0.31}
\definecolor{darkslate}{rgb}{0.28,0.24,0.55}
\definecolor{darkturquoise}{rgb}{0.00,0.81,0.82}
\definecolor{darkviolet}{rgb}{0.58,0.00,0.83}
\definecolor{deeppink}{rgb}{1.00,0.08,0.58}
\definecolor{deepsky}{rgb}{0.00,0.75,1.00}
\definecolor{dimgray}{rgb}{0.41,0.41,0.41}
\definecolor{dimgrey}{rgb}{0.41,0.41,0.41}
\definecolor{dodgerblue}{rgb}{0.12,0.56,1.00}
\definecolor{firebrick1}{rgb}{1.00,0.19,0.19}
\definecolor{firebrick2}{rgb}{0.93,0.17,0.17}
\definecolor{firebrick3}{rgb}{0.80,0.15,0.15}
\definecolor{firebrick4}{rgb}{0.55,0.10,0.10}
\definecolor{firebrick}{rgb}{0.70,0.13,0.13}
\definecolor{floralwhite}{rgb}{1.00,0.98,0.94}
\definecolor{forestgreen}{rgb}{0.13,0.55,0.13}
\definecolor{gainsboro}{rgb}{0.86,0.86,0.86}
\definecolor{ghostwhite}{rgb}{0.97,0.97,1.00}
\definecolor{gold1}{rgb}{1.00,0.84,0.00}
\definecolor{gold2}{rgb}{0.93,0.79,0.00}
\definecolor{gold3}{rgb}{0.80,0.68,0.00}
\definecolor{gold4}{rgb}{0.55,0.46,0.00}
\definecolor{goldenrod1}{rgb}{1.00,0.76,0.15}
\definecolor{goldenrod2}{rgb}{0.93,0.71,0.13}
\definecolor{goldenrod3}{rgb}{0.80,0.61,0.11}
\definecolor{goldenrod4}{rgb}{0.55,0.41,0.08}
\definecolor{goldenrod}{rgb}{0.85,0.65,0.13}
\definecolor{gold}{rgb}{1.00,0.84,0.00}
\definecolor{gray0}{rgb}{0.00,0.00,0.00}
\definecolor{gray100}{rgb}{1.00,1.00,1.00}
\definecolor{gray10}{rgb}{0.10,0.10,0.10}
\definecolor{gray11}{rgb}{0.11,0.11,0.11}
\definecolor{gray12}{rgb}{0.12,0.12,0.12}
\definecolor{gray13}{rgb}{0.13,0.13,0.13}
\definecolor{gray14}{rgb}{0.14,0.14,0.14}
\definecolor{gray15}{rgb}{0.15,0.15,0.15}
\definecolor{gray16}{rgb}{0.16,0.16,0.16}
\definecolor{gray17}{rgb}{0.17,0.17,0.17}
\definecolor{gray18}{rgb}{0.18,0.18,0.18}
\definecolor{gray19}{rgb}{0.19,0.19,0.19}
\definecolor{gray1}{rgb}{0.01,0.01,0.01}
\definecolor{gray20}{rgb}{0.20,0.20,0.20}
\definecolor{gray21}{rgb}{0.21,0.21,0.21}
\definecolor{gray22}{rgb}{0.22,0.22,0.22}
\definecolor{gray23}{rgb}{0.23,0.23,0.23}
\definecolor{gray24}{rgb}{0.24,0.24,0.24}
\definecolor{gray25}{rgb}{0.25,0.25,0.25}
\definecolor{gray26}{rgb}{0.26,0.26,0.26}
\definecolor{gray27}{rgb}{0.27,0.27,0.27}
\definecolor{gray28}{rgb}{0.28,0.28,0.28}
\definecolor{gray29}{rgb}{0.29,0.29,0.29}
\definecolor{gray2}{rgb}{0.02,0.02,0.02}
\definecolor{gray30}{rgb}{0.30,0.30,0.30}
\definecolor{gray31}{rgb}{0.31,0.31,0.31}
\definecolor{gray32}{rgb}{0.32,0.32,0.32}
\definecolor{gray33}{rgb}{0.33,0.33,0.33}
\definecolor{gray34}{rgb}{0.34,0.34,0.34}
\definecolor{gray35}{rgb}{0.35,0.35,0.35}
\definecolor{gray36}{rgb}{0.36,0.36,0.36}
\definecolor{gray37}{rgb}{0.37,0.37,0.37}
\definecolor{gray38}{rgb}{0.38,0.38,0.38}
\definecolor{gray39}{rgb}{0.39,0.39,0.39}
\definecolor{gray3}{rgb}{0.03,0.03,0.03}
\definecolor{gray40}{rgb}{0.40,0.40,0.40}
\definecolor{gray41}{rgb}{0.41,0.41,0.41}
\definecolor{gray42}{rgb}{0.42,0.42,0.42}
\definecolor{gray43}{rgb}{0.43,0.43,0.43}
\definecolor{gray44}{rgb}{0.44,0.44,0.44}
\definecolor{gray45}{rgb}{0.45,0.45,0.45}
\definecolor{gray46}{rgb}{0.46,0.46,0.46}
\definecolor{gray47}{rgb}{0.47,0.47,0.47}
\definecolor{gray48}{rgb}{0.48,0.48,0.48}
\definecolor{gray49}{rgb}{0.49,0.49,0.49}
\definecolor{gray4}{rgb}{0.04,0.04,0.04}
\definecolor{gray50}{rgb}{0.50,0.50,0.50}
\definecolor{gray51}{rgb}{0.51,0.51,0.51}
\definecolor{gray52}{rgb}{0.52,0.52,0.52}
\definecolor{gray53}{rgb}{0.53,0.53,0.53}
\definecolor{gray54}{rgb}{0.54,0.54,0.54}
\definecolor{gray55}{rgb}{0.55,0.55,0.55}
\definecolor{gray56}{rgb}{0.56,0.56,0.56}
\definecolor{gray57}{rgb}{0.57,0.57,0.57}
\definecolor{gray58}{rgb}{0.58,0.58,0.58}
\definecolor{gray59}{rgb}{0.59,0.59,0.59}
\definecolor{gray5}{rgb}{0.05,0.05,0.05}
\definecolor{gray60}{rgb}{0.60,0.60,0.60}
\definecolor{gray61}{rgb}{0.61,0.61,0.61}
\definecolor{gray62}{rgb}{0.62,0.62,0.62}
\definecolor{gray63}{rgb}{0.63,0.63,0.63}
\definecolor{gray64}{rgb}{0.64,0.64,0.64}
\definecolor{gray65}{rgb}{0.65,0.65,0.65}
\definecolor{gray66}{rgb}{0.66,0.66,0.66}
\definecolor{gray67}{rgb}{0.67,0.67,0.67}
\definecolor{gray68}{rgb}{0.68,0.68,0.68}
\definecolor{gray69}{rgb}{0.69,0.69,0.69}
\definecolor{gray6}{rgb}{0.06,0.06,0.06}
\definecolor{gray70}{rgb}{0.70,0.70,0.70}
\definecolor{gray71}{rgb}{0.71,0.71,0.71}
\definecolor{gray72}{rgb}{0.72,0.72,0.72}
\definecolor{gray73}{rgb}{0.73,0.73,0.73}
\definecolor{gray74}{rgb}{0.74,0.74,0.74}
\definecolor{gray75}{rgb}{0.75,0.75,0.75}
\definecolor{gray76}{rgb}{0.76,0.76,0.76}
\definecolor{gray77}{rgb}{0.77,0.77,0.77}
\definecolor{gray78}{rgb}{0.78,0.78,0.78}
\definecolor{gray79}{rgb}{0.79,0.79,0.79}
\definecolor{gray7}{rgb}{0.07,0.07,0.07}
\definecolor{gray80}{rgb}{0.80,0.80,0.80}
\definecolor{gray81}{rgb}{0.81,0.81,0.81}
\definecolor{gray82}{rgb}{0.82,0.82,0.82}
\definecolor{gray83}{rgb}{0.83,0.83,0.83}
\definecolor{gray84}{rgb}{0.84,0.84,0.84}
\definecolor{gray85}{rgb}{0.85,0.85,0.85}
\definecolor{gray86}{rgb}{0.86,0.86,0.86}
\definecolor{gray87}{rgb}{0.87,0.87,0.87}
\definecolor{gray88}{rgb}{0.88,0.88,0.88}
\definecolor{gray89}{rgb}{0.89,0.89,0.89}
\definecolor{gray8}{rgb}{0.08,0.08,0.08}
\definecolor{gray90}{rgb}{0.90,0.90,0.90}
\definecolor{gray91}{rgb}{0.91,0.91,0.91}
\definecolor{gray92}{rgb}{0.92,0.92,0.92}
\definecolor{gray93}{rgb}{0.93,0.93,0.93}
\definecolor{gray94}{rgb}{0.94,0.94,0.94}
\definecolor{gray95}{rgb}{0.95,0.95,0.95}
\definecolor{gray96}{rgb}{0.96,0.96,0.96}
\definecolor{gray97}{rgb}{0.97,0.97,0.97}
\definecolor{gray98}{rgb}{0.98,0.98,0.98}
\definecolor{gray99}{rgb}{0.99,0.99,0.99}
\definecolor{gray9}{rgb}{0.09,0.09,0.09}
\definecolor{gray}{rgb}{0.75,0.75,0.75}
\definecolor{green1}{rgb}{0.00,1.00,0.00}
\definecolor{green2}{rgb}{0.00,0.93,0.00}
\definecolor{green3}{rgb}{0.00,0.80,0.00}
\definecolor{green4}{rgb}{0.00,0.55,0.00}
\definecolor{greenyellow}{rgb}{0.68,1.00,0.18}
\definecolor{green}{rgb}{0.00,1.00,0.00}
\definecolor{grey0}{rgb}{0.00,0.00,0.00}
\definecolor{grey100}{rgb}{1.00,1.00,1.00}
\definecolor{grey10}{rgb}{0.10,0.10,0.10}
\definecolor{grey11}{rgb}{0.11,0.11,0.11}
\definecolor{grey12}{rgb}{0.12,0.12,0.12}
\definecolor{grey13}{rgb}{0.13,0.13,0.13}
\definecolor{grey14}{rgb}{0.14,0.14,0.14}
\definecolor{grey15}{rgb}{0.15,0.15,0.15}
\definecolor{grey16}{rgb}{0.16,0.16,0.16}
\definecolor{grey17}{rgb}{0.17,0.17,0.17}
\definecolor{grey18}{rgb}{0.18,0.18,0.18}
\definecolor{grey19}{rgb}{0.19,0.19,0.19}
\definecolor{grey1}{rgb}{0.01,0.01,0.01}
\definecolor{grey20}{rgb}{0.20,0.20,0.20}
\definecolor{grey21}{rgb}{0.21,0.21,0.21}
\definecolor{grey22}{rgb}{0.22,0.22,0.22}
\definecolor{grey23}{rgb}{0.23,0.23,0.23}
\definecolor{grey24}{rgb}{0.24,0.24,0.24}
\definecolor{grey25}{rgb}{0.25,0.25,0.25}
\definecolor{grey26}{rgb}{0.26,0.26,0.26}
\definecolor{grey27}{rgb}{0.27,0.27,0.27}
\definecolor{grey28}{rgb}{0.28,0.28,0.28}
\definecolor{grey29}{rgb}{0.29,0.29,0.29}
\definecolor{grey2}{rgb}{0.02,0.02,0.02}
\definecolor{grey30}{rgb}{0.30,0.30,0.30}
\definecolor{grey31}{rgb}{0.31,0.31,0.31}
\definecolor{grey32}{rgb}{0.32,0.32,0.32}
\definecolor{grey33}{rgb}{0.33,0.33,0.33}
\definecolor{grey34}{rgb}{0.34,0.34,0.34}
\definecolor{grey35}{rgb}{0.35,0.35,0.35}
\definecolor{grey36}{rgb}{0.36,0.36,0.36}
\definecolor{grey37}{rgb}{0.37,0.37,0.37}
\definecolor{grey38}{rgb}{0.38,0.38,0.38}
\definecolor{grey39}{rgb}{0.39,0.39,0.39}
\definecolor{grey3}{rgb}{0.03,0.03,0.03}
\definecolor{grey40}{rgb}{0.40,0.40,0.40}
\definecolor{grey41}{rgb}{0.41,0.41,0.41}
\definecolor{grey42}{rgb}{0.42,0.42,0.42}
\definecolor{grey43}{rgb}{0.43,0.43,0.43}
\definecolor{grey44}{rgb}{0.44,0.44,0.44}
\definecolor{grey45}{rgb}{0.45,0.45,0.45}
\definecolor{grey46}{rgb}{0.46,0.46,0.46}
\definecolor{grey47}{rgb}{0.47,0.47,0.47}
\definecolor{grey48}{rgb}{0.48,0.48,0.48}
\definecolor{grey49}{rgb}{0.49,0.49,0.49}
\definecolor{grey4}{rgb}{0.04,0.04,0.04}
\definecolor{grey50}{rgb}{0.50,0.50,0.50}
\definecolor{grey51}{rgb}{0.51,0.51,0.51}
\definecolor{grey52}{rgb}{0.52,0.52,0.52}
\definecolor{grey53}{rgb}{0.53,0.53,0.53}
\definecolor{grey54}{rgb}{0.54,0.54,0.54}
\definecolor{grey55}{rgb}{0.55,0.55,0.55}
\definecolor{grey56}{rgb}{0.56,0.56,0.56}
\definecolor{grey57}{rgb}{0.57,0.57,0.57}
\definecolor{grey58}{rgb}{0.58,0.58,0.58}
\definecolor{grey59}{rgb}{0.59,0.59,0.59}
\definecolor{grey5}{rgb}{0.05,0.05,0.05}
\definecolor{grey60}{rgb}{0.60,0.60,0.60}
\definecolor{grey61}{rgb}{0.61,0.61,0.61}
\definecolor{grey62}{rgb}{0.62,0.62,0.62}
\definecolor{grey63}{rgb}{0.63,0.63,0.63}
\definecolor{grey64}{rgb}{0.64,0.64,0.64}
\definecolor{grey65}{rgb}{0.65,0.65,0.65}
\definecolor{grey66}{rgb}{0.66,0.66,0.66}
\definecolor{grey67}{rgb}{0.67,0.67,0.67}
\definecolor{grey68}{rgb}{0.68,0.68,0.68}
\definecolor{grey69}{rgb}{0.69,0.69,0.69}
\definecolor{grey6}{rgb}{0.06,0.06,0.06}
\definecolor{grey70}{rgb}{0.70,0.70,0.70}
\definecolor{grey71}{rgb}{0.71,0.71,0.71}
\definecolor{grey72}{rgb}{0.72,0.72,0.72}
\definecolor{grey73}{rgb}{0.73,0.73,0.73}
\definecolor{grey74}{rgb}{0.74,0.74,0.74}
\definecolor{grey75}{rgb}{0.75,0.75,0.75}
\definecolor{grey76}{rgb}{0.76,0.76,0.76}
\definecolor{grey77}{rgb}{0.77,0.77,0.77}
\definecolor{grey78}{rgb}{0.78,0.78,0.78}
\definecolor{grey79}{rgb}{0.79,0.79,0.79}
\definecolor{grey7}{rgb}{0.07,0.07,0.07}
\definecolor{grey80}{rgb}{0.80,0.80,0.80}
\definecolor{grey81}{rgb}{0.81,0.81,0.81}
\definecolor{grey82}{rgb}{0.82,0.82,0.82}
\definecolor{grey83}{rgb}{0.83,0.83,0.83}
\definecolor{grey84}{rgb}{0.84,0.84,0.84}
\definecolor{grey85}{rgb}{0.85,0.85,0.85}
\definecolor{grey86}{rgb}{0.86,0.86,0.86}
\definecolor{grey87}{rgb}{0.87,0.87,0.87}
\definecolor{grey88}{rgb}{0.88,0.88,0.88}
\definecolor{grey89}{rgb}{0.89,0.89,0.89}
\definecolor{grey8}{rgb}{0.08,0.08,0.08}
\definecolor{grey90}{rgb}{0.90,0.90,0.90}
\definecolor{grey91}{rgb}{0.91,0.91,0.91}
\definecolor{grey92}{rgb}{0.92,0.92,0.92}
\definecolor{grey93}{rgb}{0.93,0.93,0.93}
\definecolor{grey94}{rgb}{0.94,0.94,0.94}
\definecolor{grey95}{rgb}{0.95,0.95,0.95}
\definecolor{grey96}{rgb}{0.96,0.96,0.96}
\definecolor{grey97}{rgb}{0.97,0.97,0.97}
\definecolor{grey98}{rgb}{0.98,0.98,0.98}
\definecolor{grey99}{rgb}{0.99,0.99,0.99}
\definecolor{grey9}{rgb}{0.09,0.09,0.09}
\definecolor{grey}{rgb}{0.75,0.75,0.75}
\definecolor{honeydew1}{rgb}{0.94,1.00,0.94}
\definecolor{honeydew2}{rgb}{0.88,0.93,0.88}
\definecolor{honeydew3}{rgb}{0.76,0.80,0.76}
\definecolor{honeydew4}{rgb}{0.51,0.55,0.51}
\definecolor{honeydew}{rgb}{0.94,1.00,0.94}
\definecolor{hotpink}{rgb}{1.00,0.41,0.71}
\definecolor{indianred}{rgb}{0.80,0.36,0.36}
\definecolor{ivory1}{rgb}{1.00,1.00,0.94}
\definecolor{ivory2}{rgb}{0.93,0.93,0.88}
\definecolor{ivory3}{rgb}{0.80,0.80,0.76}
\definecolor{ivory4}{rgb}{0.55,0.55,0.51}
\definecolor{ivory}{rgb}{1.00,1.00,0.94}
\definecolor{khaki1}{rgb}{1.00,0.96,0.56}
\definecolor{khaki2}{rgb}{0.93,0.90,0.52}
\definecolor{khaki3}{rgb}{0.80,0.78,0.45}
\definecolor{khaki4}{rgb}{0.55,0.53,0.31}
\definecolor{khaki}{rgb}{0.94,0.90,0.55}
\definecolor{lavenderblush}{rgb}{1.00,0.94,0.96}
\definecolor{lavender}{rgb}{0.90,0.90,0.98}
\definecolor{lawngreen}{rgb}{0.49,0.99,0.00}
\definecolor{lemonchiffon}{rgb}{1.00,0.98,0.80}
\definecolor{lightblue}{rgb}{0.68,0.85,0.90}
\definecolor{lightcoral}{rgb}{0.94,0.50,0.50}
\definecolor{lightcyan}{rgb}{0.88,1.00,1.00}
\definecolor{lightgoldenrod}{rgb}{0.93,0.87,0.51}
\definecolor{lightgoldenrod}{rgb}{0.98,0.98,0.82}
\definecolor{lightgray}{rgb}{0.83,0.83,0.83}
\definecolor{lightgreen}{rgb}{0.56,0.93,0.56}
\definecolor{lightgrey}{rgb}{0.83,0.83,0.83}
\definecolor{lightpink}{rgb}{1.00,0.71,0.76}
\definecolor{lightsalmon}{rgb}{1.00,0.63,0.48}
\definecolor{lightsea}{rgb}{0.13,0.70,0.67}
\definecolor{lightsky}{rgb}{0.53,0.81,0.98}
\definecolor{lightslate}{rgb}{0.47,0.53,0.60}
\definecolor{lightslate}{rgb}{0.47,0.53,0.60}
\definecolor{lightslate}{rgb}{0.52,0.44,1.00}
\definecolor{lightsteel}{rgb}{0.69,0.77,0.87}
\definecolor{lightyellow}{rgb}{1.00,1.00,0.88}
\definecolor{limegreen}{rgb}{0.20,0.80,0.20}
\definecolor{linen}{rgb}{0.98,0.94,0.90}
\definecolor{magenta1}{rgb}{1.00,0.00,1.00}
\definecolor{magenta2}{rgb}{0.93,0.00,0.93}
\definecolor{magenta3}{rgb}{0.80,0.00,0.80}
\definecolor{magenta4}{rgb}{0.55,0.00,0.55}
\definecolor{magenta}{rgb}{1.00,0.00,1.00}
\definecolor{maroon1}{rgb}{1.00,0.20,0.70}
\definecolor{maroon2}{rgb}{0.93,0.19,0.65}
\definecolor{maroon3}{rgb}{0.80,0.16,0.56}
\definecolor{maroon4}{rgb}{0.55,0.11,0.38}
\definecolor{maroon}{rgb}{0.69,0.19,0.38}
\definecolor{mediumaquamarine}{rgb}{0.40,0.80,0.67}
\definecolor{mediumblue}{rgb}{0.00,0.00,0.80}
\definecolor{mediumorchid}{rgb}{0.73,0.33,0.83}
\definecolor{mediumpurple}{rgb}{0.58,0.44,0.86}
\definecolor{mediumsea}{rgb}{0.24,0.70,0.44}
\definecolor{mediumslate}{rgb}{0.48,0.41,0.93}
\definecolor{mediumspring}{rgb}{0.00,0.98,0.60}
\definecolor{mediumturquoise}{rgb}{0.28,0.82,0.80}
\definecolor{mediumviolet}{rgb}{0.78,0.08,0.52}
\definecolor{midnightblue}{rgb}{0.10,0.10,0.44}
\definecolor{mintcream}{rgb}{0.96,1.00,0.98}
\definecolor{mistyrose}{rgb}{1.00,0.89,0.88}
\definecolor{moccasin}{rgb}{1.00,0.89,0.71}
\definecolor{navajowhite}{rgb}{1.00,0.87,0.68}
\definecolor{navyblue}{rgb}{0.00,0.00,0.50}
\definecolor{navy}{rgb}{0.00,0.00,0.50}
\definecolor{oldlace}{rgb}{0.99,0.96,0.90}
\definecolor{olivedrab}{rgb}{0.42,0.56,0.14}
\definecolor{orange1}{rgb}{1.00,0.65,0.00}
\definecolor{orange2}{rgb}{0.93,0.60,0.00}
\definecolor{orange3}{rgb}{0.80,0.52,0.00}
\definecolor{orange4}{rgb}{0.55,0.35,0.00}
\definecolor{orangered}{rgb}{1.00,0.27,0.00}
\definecolor{orange}{rgb}{1.00,0.65,0.00}
\definecolor{orchid1}{rgb}{1.00,0.51,0.98}
\definecolor{orchid2}{rgb}{0.93,0.48,0.91}
\definecolor{orchid3}{rgb}{0.80,0.41,0.79}
\definecolor{orchid4}{rgb}{0.55,0.28,0.54}
\definecolor{orchid}{rgb}{0.85,0.44,0.84}
\definecolor{palegoldenrod}{rgb}{0.93,0.91,0.67}
\definecolor{palegreen}{rgb}{0.60,0.98,0.60}
\definecolor{paleturquoise}{rgb}{0.69,0.93,0.93}
\definecolor{paleviolet}{rgb}{0.86,0.44,0.58}
\definecolor{papayawhip}{rgb}{1.00,0.94,0.84}
\definecolor{peachpuff}{rgb}{1.00,0.85,0.73}
\definecolor{peru}{rgb}{0.80,0.52,0.25}
\definecolor{pink1}{rgb}{1.00,0.71,0.77}
\definecolor{pink2}{rgb}{0.93,0.66,0.72}
\definecolor{pink3}{rgb}{0.80,0.57,0.62}
\definecolor{pink4}{rgb}{0.55,0.39,0.42}
\definecolor{pink}{rgb}{1.00,0.75,0.80}
\definecolor{plum1}{rgb}{1.00,0.73,1.00}
\definecolor{plum2}{rgb}{0.93,0.68,0.93}
\definecolor{plum3}{rgb}{0.80,0.59,0.80}
\definecolor{plum4}{rgb}{0.55,0.40,0.55}
\definecolor{plum}{rgb}{0.87,0.63,0.87}
\definecolor{powderblue}{rgb}{0.69,0.88,0.90}
\definecolor{purple1}{rgb}{0.61,0.19,1.00}
\definecolor{purple2}{rgb}{0.57,0.17,0.93}
\definecolor{purple3}{rgb}{0.49,0.15,0.80}
\definecolor{purple4}{rgb}{0.33,0.10,0.55}
\definecolor{purple}{rgb}{0.63,0.13,0.94}
\definecolor{red1}{rgb}{1.00,0.00,0.00}
\definecolor{red2}{rgb}{0.93,0.00,0.00}
\definecolor{red3}{rgb}{0.80,0.00,0.00}
\definecolor{red4}{rgb}{0.55,0.00,0.00}
\definecolor{red}{rgb}{1.00,0.00,0.00}
\definecolor{rosybrown}{rgb}{0.74,0.56,0.56}
\definecolor{royalblue}{rgb}{0.25,0.41,0.88}
\definecolor{saddlebrown}{rgb}{0.55,0.27,0.07}
\definecolor{salmon1}{rgb}{1.00,0.55,0.41}
\definecolor{salmon2}{rgb}{0.93,0.51,0.38}
\definecolor{salmon3}{rgb}{0.80,0.44,0.33}
\definecolor{salmon4}{rgb}{0.55,0.30,0.22}
\definecolor{salmon}{rgb}{0.98,0.50,0.45}
\definecolor{sandybrown}{rgb}{0.96,0.64,0.38}
\definecolor{seagreen}{rgb}{0.18,0.55,0.34}
\definecolor{seashell1}{rgb}{1.00,0.96,0.93}
\definecolor{seashell2}{rgb}{0.93,0.90,0.87}
\definecolor{seashell3}{rgb}{0.80,0.77,0.75}
\definecolor{seashell4}{rgb}{0.55,0.53,0.51}
\definecolor{seashell}{rgb}{1.00,0.96,0.93}
\definecolor{sienna1}{rgb}{1.00,0.51,0.28}
\definecolor{sienna2}{rgb}{0.93,0.47,0.26}
\definecolor{sienna3}{rgb}{0.80,0.41,0.22}
\definecolor{sienna4}{rgb}{0.55,0.28,0.15}
\definecolor{sienna}{rgb}{0.63,0.32,0.18}
\definecolor{skyblue}{rgb}{0.53,0.81,0.92}
\definecolor{slateblue}{rgb}{0.42,0.35,0.80}
\definecolor{slategray}{rgb}{0.44,0.50,0.56}
\definecolor{slategrey}{rgb}{0.44,0.50,0.56}
\definecolor{snow1}{rgb}{1.00,0.98,0.98}
\definecolor{snow2}{rgb}{0.93,0.91,0.91}
\definecolor{snow3}{rgb}{0.80,0.79,0.79}
\definecolor{snow4}{rgb}{0.55,0.54,0.54}
\definecolor{snow}{rgb}{1.00,0.98,0.98}
\definecolor{springgreen}{rgb}{0.00,1.00,0.50}
\definecolor{steelblue}{rgb}{0.27,0.51,0.71}
\definecolor{tan1}{rgb}{1.00,0.65,0.31}
\definecolor{tan2}{rgb}{0.93,0.60,0.29}
\definecolor{tan3}{rgb}{0.80,0.52,0.25}
\definecolor{tan4}{rgb}{0.55,0.35,0.17}
\definecolor{tan}{rgb}{0.82,0.71,0.55}
\definecolor{thistle1}{rgb}{1.00,0.88,1.00}
\definecolor{thistle2}{rgb}{0.93,0.82,0.93}
\definecolor{thistle3}{rgb}{0.80,0.71,0.80}
\definecolor{thistle4}{rgb}{0.55,0.48,0.55}
\definecolor{thistle}{rgb}{0.85,0.75,0.85}
\definecolor{tomato1}{rgb}{1.00,0.39,0.28}
\definecolor{tomato2}{rgb}{0.93,0.36,0.26}
\definecolor{tomato3}{rgb}{0.80,0.31,0.22}
\definecolor{tomato4}{rgb}{0.55,0.21,0.15}
\definecolor{tomato}{rgb}{1.00,0.39,0.28}
\definecolor{turquoise1}{rgb}{0.00,0.96,1.00}
\definecolor{turquoise2}{rgb}{0.00,0.90,0.93}
\definecolor{turquoise3}{rgb}{0.00,0.77,0.80}
\definecolor{turquoise4}{rgb}{0.00,0.53,0.55}
\definecolor{turquoise}{rgb}{0.25,0.88,0.82}
\definecolor{violetred}{rgb}{0.82,0.13,0.56}
\definecolor{violet}{rgb}{0.93,0.51,0.93}
\definecolor{wheat1}{rgb}{1.00,0.91,0.73}
\definecolor{wheat2}{rgb}{0.93,0.85,0.68}
\definecolor{wheat3}{rgb}{0.80,0.73,0.59}
\definecolor{wheat4}{rgb}{0.55,0.49,0.40}
\definecolor{wheat}{rgb}{0.96,0.87,0.70}
\definecolor{whitesmoke}{rgb}{0.96,0.96,0.96}
\definecolor{white}{rgb}{1.00,1.00,1.00}
\definecolor{yellow1}{rgb}{1.00,1.00,0.00}
\definecolor{yellow2}{rgb}{0.93,0.93,0.00}
\definecolor{yellow3}{rgb}{0.80,0.80,0.00}
\definecolor{yellow4}{rgb}{0.55,0.55,0.00}
\definecolor{yellowgreen}{rgb}{0.60,0.80,0.20}
\definecolor{yellow}{rgb}{1.00,1.00,0.00}

\newcommand{\mc}{\multicolumn}

\usepackage{graphicx}
\usepackage{times} 
\usepackage{amssymb}
\usepackage{amsmath}
\usepackage{lscape}
\usepackage{url}
\newif\ifAMStwofonts
\AMStwofontstrue

\def\apj{ApJ}
\def\mnras{MNRAS}

\def\araa{ARA\&A}                
\def\aap{A\&A}                   
\def\aj{AJ}                      
\def\apjs{ApJS}                  
\def\apjl{ApJ}                   

\def\kpc{\hbox{$\rm\thinspace kpc$}}

\def\pcmsq{\hbox{$\rm\thinspace cm^{-2}$}}
\def\kmpspmpc{\hbox{$\rm\thinspace km~s^{-1}~Mpc^{-1}$}}

\def\kev{\hbox{$\rm\thinspace keV$}}

\def\ergpcmsqps{\hbox{$\rm\thinspace erg~cm^{-2}~s^{-1}$}}

\def\ergps{\hbox{$\rm\thinspace erg~s^{-1}$}}

\begin{document}

\title[The X-ray and radio lobes of 4C23.56] {The X-ray and radio-emitting plasma lobes 
of 4C23.56: further evidence of recurrent jet activity and high acceleration energies}
\author[Blundell \& Fabian]
{\parbox[]{6.in} {Katherine~M.\,Blundell$^{1}$ and A.C.\,Fabian$^{2}$}\\\\
  \footnotesize
 $^{1}$University of Oxford, Astrophysics, Keble Road, Oxford OX1 3RH\\
 $^{2}$Institute of Astronomy, Madingley Road, Cambridge CB3 0HA\\
}
\maketitle

\begin{abstract}
  New Chandra observations of the giant ($\sim 0.5$\,Mpc) radio galaxy
  4C23.56 at $z = 2.5$ show X-rays in a linear structure aligned with
  its radio emission, but seemingly anti-correlated with the detailed
  radio structure.  Consistent with the powerful, high-$z$ giant radio
  galaxies we have studied previously, X-rays seem to be invariably
  found where the lobe plasma is oldest even where the radio emission
  has long since faded.  The hotspot complexes seem to show structures
  resembling the double shock structure exhibited by the largest radio
  quasar 4C74.26, with the X-ray shock again being offset closer to the
  nucleus than the radio synchrotron shock.  In the current paper, the
  offsets between these shocks are even larger, at $\sim$35\,kpc.
  Unusually for a classical double (FRII) radio source, there is
  smooth low surface-brightness radio emission associated with the
  regions {\it beyond} the hotspots (i.e.\ further away from the
  nucleus than the hotspots themselves), which seems to be symmetric
  for the ends of both jets.  We consider possible explanations for
  this phenomenon, and reach the conclusion that it arises from
  high-energy electrons, recently accelerated in the nearby radio
  hotspots that are leaking into a pre-existing weakly-magnetized
  plasma that are symmetric relic lobes fed from a previous episode of
  jet activity.  This contrasts with other manifestations of previous
  epochs of jet ejection in various examples of classical double radio
  sources namely (i) double-double radio galaxies by e.g. Schoenmakers
  et al, (ii) the double-double X-ray/radio galaxies by Laskar et al
  and (iii) the presence of a relic X-ray counter-jet in the
  prototypical classical double radio galaxy, Cygnus\,A by Steenbrugge
  et al.  The occurrence, still more the prevalence, of multi-episodic
  jet activity in powerful radio galaxies and quasars indicates that
  they may have a longer lasting influence on the on-going structure
  formation processes in their environs than previously presumed.
\end{abstract}

\begin{keywords}
galaxies: high-redshift, galaxies: jets, X-rays: individual (4C23.56)
\end{keywords}

\section{Introduction}
X-ray studies of classical double \citep[FR\,II;][]{Fanaroff1974}
radio galaxies and quasars seem --- at significant distances out from
their nuclei --- to display two separate phenomena besides their
compact, active galactic nuclei: the first of these is that
considerable {\it extended} X-ray emission may be associated with the
oldest parts of the lobes \citep[e.g.\
][]{Blundell2006,Johnson2007,Erlund2008a,Laskar2010}.  This is
exemplified by our X-ray study of the powerful, giant, $z \sim 2$
radio galaxy 6C0905+39: our Chandra observations \citep{Blundell2006}
revealed that the {\it oldest} parts of the lobes (nearest to the
nucleus, yet distinct from it) were predominantly where fairly smooth
X-ray emission was detected even though the radio emission from that
plasma (accelerated $>10^8$\,years ago when the hotspots were that
close to the nucleus) had faded.  This X-ray emission, extended over
several 100\,kpc, has been confirmed by subsequent, deeper XMM
observations \citep{Erlund2008a}.  It arises via inverse-Compton
scattering of CMB photons (hereafter ICCMB) by the ``spent'' (i.e.\
relic) synchrotron particles having Lorentz factors of only $\gamma
\sim 10^3$, too low to give synchrotron radiation detectable by
current radio facilities (this requires rather higher Lorentz factors
in the typical magnetic field strengths of aged lobe plasma).  Where
we detect ICCMB, this is a tracer of the $\gamma \sim 10^3$ particles;
if these particles had been initially accelerated to higher energies
then it is likely they would have {\it previously} radiated GHz-radio
synchrotron before they lost significant amounts of their energy via
expansion losses.  An example of a relic radio source now observable
only via ICCMB emission is HDF\,130 at $z \sim 2$
\citep{Fabian2009}. As time goes by, even the ICCMB emission from
relic lobes will fade and become undetectable with present X-ray
instrumental capability \citep{Mocz2010a}.

\begin{figure*}
\includegraphics[width=17.6cm]{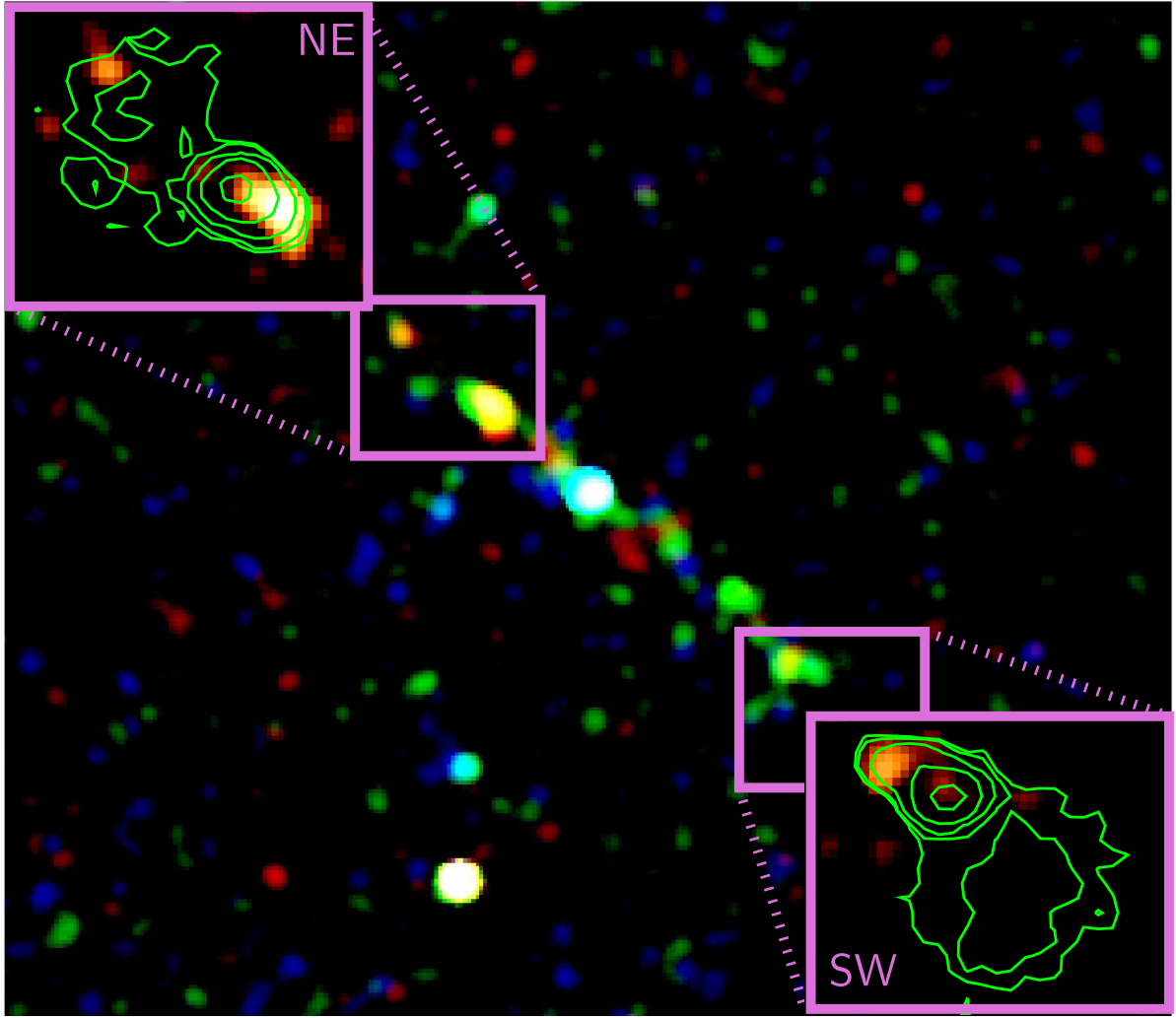}
\caption{\label{fig:overlay} Smoothed Chandra image as colourscale
  showing that ICCMB X-rays are associated with the oldest parts of
  the lobe, {\it nearest} to the central nucleus (coloured white)
  though clearly resolved from it.  A very similar picture was found
  from XMM observations \citep{Johnson2007}.  Such ICCMB emission is a
  tracer of ``spent'' synchrotron particles: too lacking in energy to
  give synchrotron radiation at the usual radio wavelengths, but
  sufficiently energetic to upscatter CMB photons to become X-ray
  photons.  As such this emission delineates the presence of $\gamma
  \sim 10^3$ particles.  Red is soft and blue is hard.  The green
  contours represent 5\,GHz radio emission.  The outermost extended
  radio regions external to the hotspots are considered in
  Sec\,\ref{sec:extended}.  The regions inside the hotspots, towards
  the nucleus are referred to as the tails; these are resolved more
  clearly from the compact hotspots in the high-resolution 8\,GHz
  image of \citet[][their fig.\,35]{Carilli1997}.}
\end{figure*}

The second phenomenon often revealed in powerful classical double
radio sources is {\it compact} X-ray emission associated, though not
co-spatial, with the radio hotspots.   Angular resolution often
presents a challenge to diagnosing the spatial separation of the
compact X-ray shock structure from the compact radio shocks, but in
the case of the giant nearby radio quasar 4C74.26, these are separated
by a projected distance of 19\,kpc \citep{Erlund2010} while in 4C19.44
the separation is   14.5\,kpc \citep{Sambruna2002}.  

The particular object under study in this paper, 4C23.56, is a
suitable target to further explore these two phenomena to try and
discern what are their fundamental characteristics.   4C23.56 is a
giant radio galaxy extending over 60\,arcminutes, corresponding to
492\,kpc projected on the plane of the Sky, previously studied in
X-rays by \citet{Johnson2007}. 

Throughout this paper, all errors are quoted at $1\sigma$ unless
otherwise stated and the cosmology is $H_{\rm 0} = 71$\kmpspmpc,
$\Omega_{0}=1$ and $\Lambda_{0} = 0.73$.  One arcsecond represents
$8.203$\kpc\ on the plane of the sky at the redshift of 4C\,23.56 of
2.48 \citep{Rottgering1997} and the Galactic absorption along the
line-of-sight towards this object is $1.29 \times 10^{21}$\pcmsq\
\citep{dickeylockman90} but reported as $9.19 \times 10^{20}$\pcmsq\
in the Leiden, Argentina, Bonn (LAB) survey \citep{Kalberla2005}.

\section{Data}

\begin{figure*}
\includegraphics[width=17.6cm]{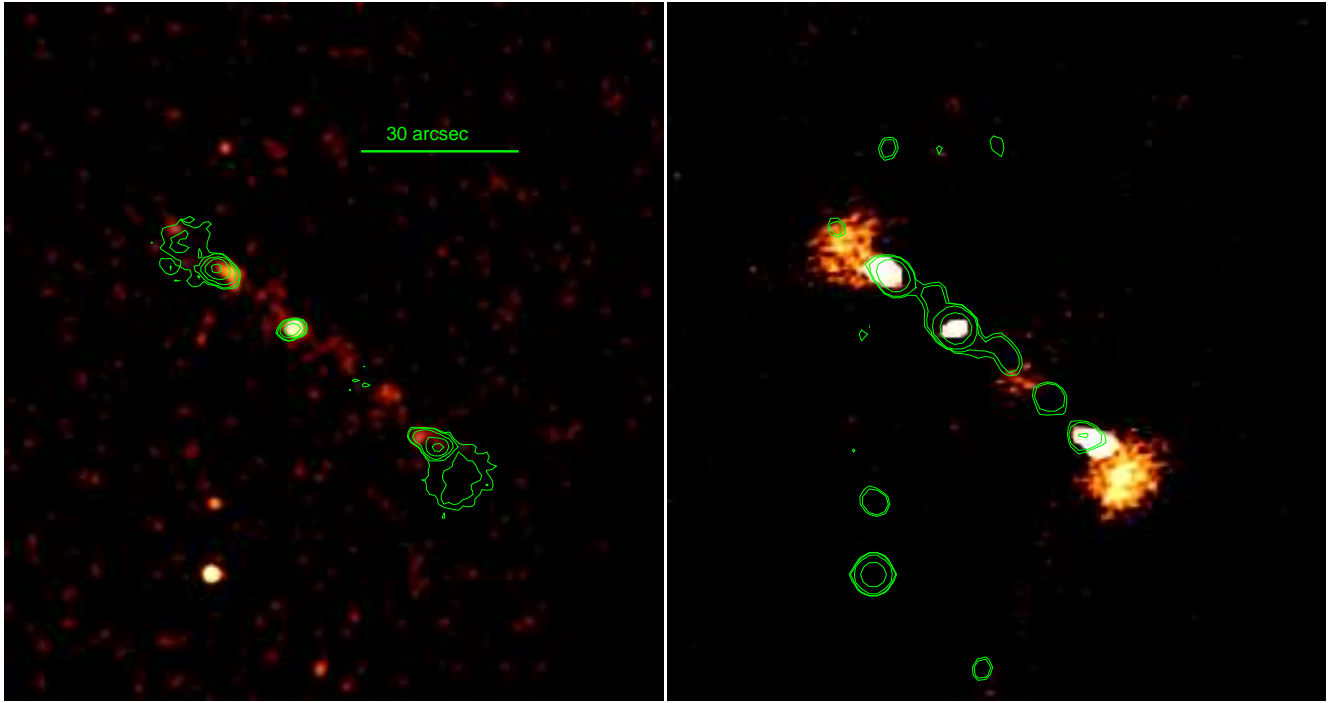}
\caption{Left: overlay of 4.8\,GHz contours on 0.5--3\,keV X-ray
  colour-scale.  Right: 0.5--3\,keV X-rays, convolved with a Gaussian
  of 2.5 arcseconds, shown as contours overlaid on 4.8\,GHz
  colourscale.  \label{fig:radioxrayoverlay}}
\end{figure*}

\subsection{New Chandra data}

4C\,23.56 was observed with Chandra ACIS-S for 91.69~ks on 2009 Aug
16--17. The data were taken in {\sc VFAINT} mode and processed accordingly to
improve background reduction. There were no significant background
flares during the observation so the whole dataset was used. After
background subtraction we detect 417 counts from the radio source in
the 0.5--7~keV band, of which 256 are from the nucleus and 161 are
from extended emission. 


\subsection{Radio data}

Data at 5\,GHz had been observed by the VLA in its fairly extended
B-configuration and its most compact D-configuration.  These were
retrieved from the archive under project codes AC234 and AR409
respectively and reduced separately using standard procedures within
the AIPS software package including self-calibration for phases only.
Confusing sources were subtracted from each UV-dataset using,
sequentially, BOX2CC, CCEDT and UVSUB.  The two datasets were
concatenated, and cross-calibration was performed on the overlapping
baselines between 0.5\,k$\lambda$ and 16\,k$\lambda$.  The extended
structure exterior to the hotspots was initially not included in any
self-calibration model.  The presence of this extended structure was
robust to taking only half of the data at a time and seems to be a
persistent and real feature of the brightness distribution.  The
north-east extended outer emission has a flux density of $8.4 \pm
0.1$\,mJy while that of the south-west extended outer emission has a
flux density of $6.4 \pm 0.1$\,mJy.

\begin{figure}
\includegraphics[width=8.5cm,angle=0]{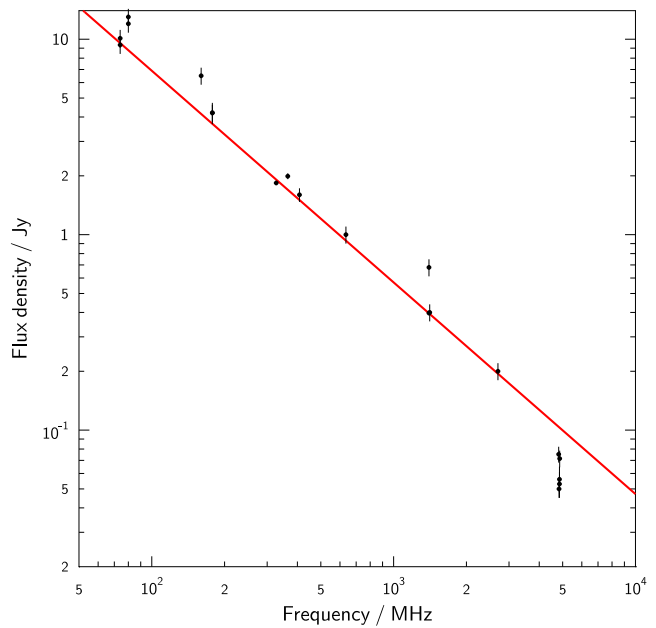}
\caption{\label{fig:intspectrum} Spectrum of 4C23.56 at radio
  wavelengths; most datapoints are taken from NED, but those at
  327\,MHz and 4.7\,GHz were taken from re-reduced radio data from the
  VLA archive.  The line shown is fitted to all points with
  frequencies below 3\,GHz and follows $\nu^{-1.1}$.  }
\end{figure}

The integrated radio spectrum is plotted in
Figure\,\ref{fig:intspectrum}.  Some of the datapoints are above the
best-fit line but these correspond to survey measurements where often
low-resolution and confusion overestimates the flux densities in what
is a fairly clustered part of the radio sky.  The radio spectrum is
consistent with being a straight power-law following $\nu^{-1.1}$ up
to frequencies of 1\,GHz.  The highest frequency points shown in the
plot, at 5\,GHz are (from top to bottom): our combined B+D
configuration data, B-configuration data and below these are various
surveys \citep{Griffith1990,Gregory1991}.  We suspect that the reason
for these particular surveys apparently underestimating the flux
density of this source compared with those obtained from pointed VLA
observations is that only the south-west lobe of this arcminute-sized
radio source is fitted in an automated manner by the surveys.  With the
extended spectral region described with a spectral index $\alpha$ =
$1.1$, there is a (minimum) $\Delta\alpha \sim 0.5$ as inferred from
the spectral index above 2.7\,GHz to the 4.8\,GHz data-point from our
VLA measurements.

From the purely monochromatic radio view, as pointed out by
\citet{Chambers1996}, the two opposite lobes and the nucleus are
perfectly co-linear although the length of the western arm is twice
that of the eastern arm.

\section{Magnetic field estimate}


An estimate of the magnetic field strength in the north-east inner
radio tail (nearer the hotspot than the nucleus, shown in the NE inset
to Fig\,\ref{fig:overlay}) may be obtained by estimating the ratio of
the co-spatial radio synchrotron luminosity with the ICCMB X-ray
luminosity \citep{Govoni2004}.  Care was taken to exclude the radio
emission from the actual compact hotspot, which has no counterpart
X-ray emission.  With a radio flux density of 5\,mJy and an X-ray flux
density of $2.17 \times 10^{-15}$\,\ergpcmsqps, the magnetic field
strength of this plasma is 16.2\,$\mu$G (a.k.a.\ 1.62 nT).  While this
is an uncertain estimate, not least because of the assumption that the
X-ray emission and radio synchrotron emission arise from the same
particle population, which as described in Sec\,\ref{sec:extended} is
doubtful, it is an entirely plausible value consistent with other
estimates of magnetic field strengths in the lobes of radio galaxies
and quasars.  In such a magnetic field strength, the particles
responsible for the synchrotron emission at 5\,GHz will have Lorentz
factors of $1.03 \times 10^4$, an order of magnitude more energetic
than those responsible for inverse-Compton scattering CMB photons into
X-ray photons.  For comparison, the magnetic field strength may be
estimated using minimum-energy assumptions \citet[e.g.\ ][]{Miley1980}
for fiducial values of the lower limit of the the energy distribution
of the synchrotron particles, $\gamma_{\rm min}$.  As discussed by
\citet{Blundell2006}, the inferred magnetic field is lower for a
higher assumed $\gamma_{\rm min}$.  Example values are: 218.2\,$\mu$G,
56.7\,$\mu$G, 28.9\,$\mu$G and 16.0\,$\mu$G for $\gamma_{\rm min} = 1,
100, 1000, 7500$ respectively.   We remind the reader that consistency
seems to emerge on the basis of higher values of $\gamma_{\rm min}$,
the resolution of the radio/X-ray overlaid image could hide the fact
that the X-ray emitting plasma and radio emitting plasma can only be
assumed to be co-spatial. 

\begin{figure}
\includegraphics[width=8.5cm,angle=0]{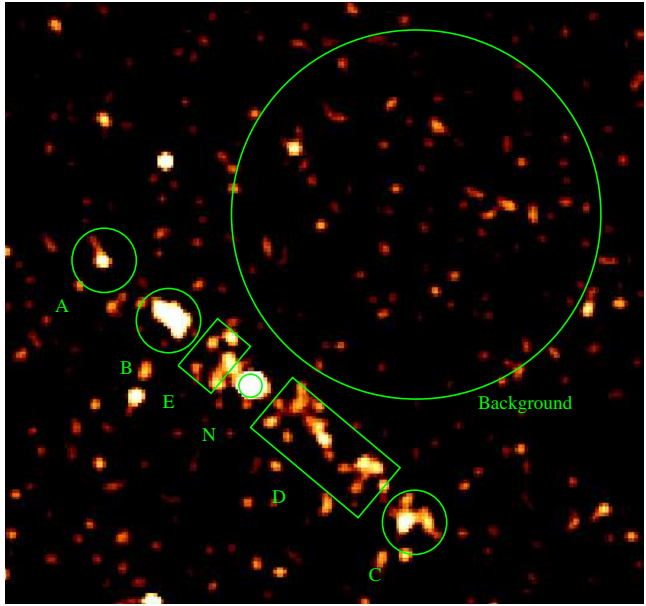}
\caption{\label{fig:regions} Illustration of the regions fitted from
  the X-ray observations to give the data presented in
  Table\,\ref{tab:regions}.}
\end{figure}

\section{Extended X-ray emission in the oldest parts of the lobes}
\label{sec:extended}

Although the X-ray counts are slightly sparse across 4C23.56,
luminosities and photon indices were obtained for regions A--E
depicted in Fig\,\ref{fig:regions} (the nucleus is discussed
separately in Sec\,\ref{sec:nucleus}).  The photon indices fitted for
the non-nuclear regions span a range of 2 -- 2.8.  This is broadly
consistent with the value inferred by \citet{Johnson2007} of $\Gamma =
2.6$ for photon index of the lobe plasma although their observations
from XMM, having lower angular resolution, would have been challenged
by contamination from the nucleus.  Although the Chandra photon
indices are consistent with those inferred from the XMM measurements,
they are inconsistent with the radio spectral index of $\alpha = 1.1$,
which would for a single continuous power-law imply a photon index of
$\Gamma = 2.1$.  The radio synchrotron electrons are thus drawn from a
harder (flatter) distribution than the electrons responsible for the
ICCMB emission.  We suggest that this points to a more recently
accelerated population of electrons as the synchrotron population,
than the older (`spent' synchrotron) lower energy electrons that
upscatter CMB photons into the X-ray bands.  This is consistent with
the picture of a previous epoch of jet activity, whose only tell-tale
sign is from relic ICCMB emission, and a current epoch of jet
activity, evinced by radio synchrotron hotspots.  We continue this
line of thought in Sec\,\ref{sec:fluff} as we consider the unusual
characteristic of the extended radio emission {\it external} to the
hotspots in this powerful radio galaxy.

The most north-easterly source along the radio axis, delineated by
Region A in Fig\,\ref{fig:regions}, has a colour, appearing as orange
in Fig\,\ref{fig:overlay} which is clearly rare amongst the background
sources.  Its position, being almost exactly along the radio axis, and
its colour resembling that of Region C suggests that it may be a part
of the non-thermal emission associated with the jet activity of this
object.  The number of counts is however, too few to permit detailed
analysis, other than (weakly) constraining its photon index or to
securely identify its role (if any) within this active galaxy's
extended emission.

\begin{table} 
\center
\begin{tabular}{cll}
\hline
Region       & Luminosity & Photon index   \\
                     & \mc{1}{c}{($\times 10^{44}$\,\ergps)}  & \\
\hline
A & $0.32 \pm $ & $1.96 \pm 0.82$ \\
B & $1.29 \pm 0.21 $ & $2.83 \pm 0.50$ \\
C & $1.09 \pm $ & $2.28 \pm 0.82$ \\
D & $1.13 \pm 0.20 $ & $2.79 \pm 0.51$ \\
E & $0.73 \pm $ & for fixed $\Gamma = 2$ \\
E & $1.1 \pm 0.3$ & for fixed $\Gamma = 2.8$ \\
N & $41.52 \pm $ & $2.27 \pm 1.05$ \\
\hline
\end{tabular}
\caption[table info]
  {Luminosities are measured between 2 and 10\,keV in the rest frame and absorption corrected (both Galactic and intrinsic).  }
\label{tab:regions} 
\end{table}

\section{Extended radio emission beyond hotspots}
\label{sec:fluff}

The insets to Fig\,\ref{fig:overlay} clearly reveal how there is
smooth radio emission {\it outside} of the hotspots, the regions of
enhanced surface brightness which, usually in classical double FRII
radio sources, are believed to be where the jets impinge and shock on
the intergalactic medium.  This outer emission is most
uncharacteristic in powerful FRII radio sources.
There appear to be no counterpart X-rays associated
  with this smooth emission, and it seems likely that X-rays would
  have been detected if they were there at the same level as the
  X-rays associated with the internal tails of the hotspots: the ratio
  of the X-ray counts in the tail region to the co-spatial, diffuse
  4.9\,GHz radio blob region (i.e.\ excluding the hotspot) as 13 to
  $<5.7$ (2\,$\sigma$ limit) i.e.\ $ >2.3$, whereas the ratio of the
  radio fluxes for the same regions is 0.5.  This means that we would
have easily detected the external regions beyond the hotspots if they
have the same electron population as the tail regions. The external
extended regions, lacking associated X-ray emission, therefore are
deficient in $\gamma \sim 1000$ electrons compared with the internal
tails.  The radio emission in these regions is presumably synchrotron
radiation and thus even the regions external to the hotspots require a
magnetic field.  Such external features are unknown in high-$z$,
powerful radio sources which suggests that IGM magnetic field
strengths are usually too low to be sufficient to ``illuminate" the
presence of high-energy electrons.  Given the symmetry of these
external extended features, we suggest that the most likely origin for
the magnetic field responsible for this emission arises from
pre-existing lobe plasma encountered by the outward moving radio
hotspots, and possibly somewhat compressed by them.  A weak magnetic
field will illuminate high-$\gamma$ electrons and if the deficiency in
$\gamma \sim 1000$ particles is indeed real then this would strongly
suggest that the turnover frequency in the energy spectrum of
particles accelerated in the hotspot complexes is very significantly
above $10^3$.  We suggest that these high-energy particles leak out
from the radio hotspot and, being ultra-relativistic, rapidly pervade
the relic, expanded plasma having a very weak, but not necessarily
tangled magnetic field.  The extent to which ultra-relativistic
electrons can pervade a pre-existing weakly-magnetized plasma may be
described by a random walk of particles across randomly oriented
regions of magnetic field \citep{Rechester1978}.  The efficacy of this
transport mechanism depends critically on the configuration (tangling)
of the magnetic field structure within the plasma which in turn
depends on its original configuration and the goemetry of any
subsequent expansion.  Anomalous diffusion mechanisms in plasma
physics are currently not fully explored, but following
\citet{Duffy1995} and assuming a tangling scale of 10\kpc\ (plausible
for expanded lobe material often 10s of kpc across in spatial extent)
then in $10^5$ years the rms distance diffused is 10s of kpc,
comparable with the extent of the external regions of plasma beyond
the hotspots, as seen in the insets of Fig\,\ref{fig:overlay}.

\section{Nucleus, and supply of fuel for a new episode of jet ejection}
\label{sec:nucleus}

\begin{figure}
\includegraphics[width=6cm,angle=270]{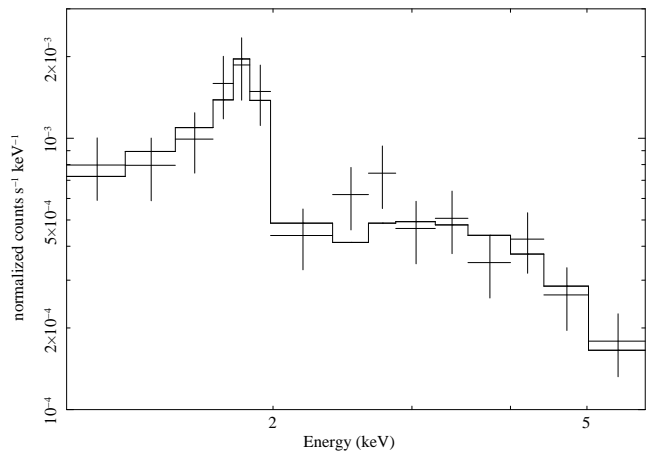}
\caption{X-ray spectrum of the nucleus of 4C23.56 showing tentative evidence for a 6.4\,kev\ iron line redshifted to 6.4/($1+z$) = 1.8\,keV, just below the deep neutral iron K edge. }
\label{fig:nuclearspectrum}
\end{figure}

The absorption-corrected luminosity of 4C\,23.56's nucleus in the
2--10\,\kev\ band is $41.5 \times 10^{44}$\,\ergps\ which for a
typical quasar nucleus \citep[e.g.][]{Elvis1994} would make the total
bolometric nuclear luminosity to be $2 \times 10^{47}$\,\ergps\ and as
such is one of the most energetic objects/nuclei in the universe.
This luminosity is very close to the Eddington-limited luminosity for
a supermassive black hole of a billion solar masses.  Although recent
research has suggested that objects with luminosities near the
Eddington-limit tend not to have radio jets
\citep[e.g.][]{Panessa2007,Sikora2007,Maoz2007} 4C23.56, just like the
other $\sim$Mpc, $z \sim 2$ radio galaxy 6C0905+32 we studied
\citep{Erlund2008b}, is another demonstrable counter-example.  We
note, however, that the Eddington limit is calculated with the
simplifying, incorrect, assumption of spherical symmetry.  The X-ray
spectrum of this nucleus, shown in Fig\,\ref{fig:nuclearspectrum}
shows tentative evidence for a redshifted 6.4\,\kev\ iron line which
is likely to be arise from the strong absorber.

The X-ray photon index of the nucleus, as listed in
Table\,\ref{tab:regions}, is fitted to be $\Gamma = 2.27 \pm 1.05$.
Though poorly constrained, it is interestingly consistent with the
high-frequency radio spectral index of the nucleus, reported between
5\,GHz and 15\,GHz by \citet{Chambers1996} to be unusually steep at
$\alpha = 1.44$.  This is in contrast with the lower radio frequency
spectral index of the nucleus of $\alpha = 0.62$.

The nucleus of this powerful radio galaxy is highly obscured, having a
column density of $0.79 \pm 0.5 \times 10^{24}$\,\pcmsq, slightly
below the Compton-thick threshold of $1.5 \times 10^{24}$\,\pcmsq\ at
which the Thomson-scattering optical depth reaches unity.

In the context of active nuclei that may be exhibiting multiple cycles
of relativistic jet activity, \citet[][and references
therein]{Salter2010} have pointed out that the presence of a higher
than average amount of neutral hydrogen absorption seems to be
associated with such ``rejuvenated" or ``re-activated" or
multi-episodic radio galaxies.  This characteristic appears to apply
to sources of all sizes from compact to giant.  Although we have no HI
observations of 4C23.56 to bring to bear against the two strands of
evidence that it has exhibited more than one episode of jet activity
([i] high-energy emission external to the hotspots requiring a
pre-existing magnetic field and [ii] the harder synchrotron spectrum
compared with that of the ICCMB emission) there are two indications of
a plentiful supply of gas in the vicinity of this nucleus: (i) the
considerable column density of $0.79 \pm 0.5 \times 10^{24}$\,\pcmsq,
(ii) the presence of two opposite cones, having $\sim 90^{\circ}$
opening angles, revealed in narrow-band Lyman-$\alpha$ imaging by
\citet{Knopp1997} and presumably illuminated by ionising radiation
from the nucleus and (iii) a rotating gas cloud with dynamical mass
inferred from Keck spectroscopy by \citet{VillarMartin2003} to be $2.9
\times 10^{12}$\,M$_{\odot}$.

We also note the remarkable structural asymmetry in this system:
  the SW arm is twice the length of the NE arm, and at half the
  distance along the SW arm there is some radio emission and an
  absence of X-ray emission (see Fig\,\ref{fig:radioxrayoverlay}).
  Whether this corresponds to evidence of alternating activity in the
  opposite jets (``flip-floping'') or to asymmetries induced by the
  presence of different plasma or other environmental differences is
  not clear with the present data, but follow-up observations will be
  sought from the eVLA. 

\section{Double shock X-ray/radio structures and particle acceleration}

Examination of the brightest peaks on the radio image, the compact
hotspot peaks (see insets in Fig\,\ref{fig:overlay}) reveals that the
brightest compact radio hotspot peaks are offset from the brightest
compact X-ray peaks in the same sense as each other: the X-ray
hotspots are offset closer to the nucleus.  This is in the same sense
as in the unusually nearby and giant radio quasar 4C74.26 studied by
\citet{Erlund2010} (the offset in this object is 19\,kpc projected on
the plane of the sky) and for the quasar 4C19.44 \citep{Sambruna2002}
which has an offset of 14.5\,kpc.  4C23.56 has a much larger offset of
$\sim 35$\,kpc; we note that the radio luminosity of 4C23.56 at
$10^{28.22}$\,W\,Hz$^{-1}$\,sr$^{-1}$ is over three orders of
magnitude higher than that of 4C74.26 at
$10^{24.93}$\,W\,Hz$^{-1}$\,sr$^{-1}$.  The origin of these offsets
are still unclear but are explored in \citet{Erlund2010}.  A
  particular possibility we explore in that paper is that the shock
  nearer to the nucleus, (the X-ray peak) is from a population of
  high-energy electrons that cool radiatively as they flow
  downstream.  A subsequent shock, further downstream, is where these
  cooled electrons are compressed and radiate in the radio. 

\section{Conclusions}

This combined X-ray and radio study demonstrates a continuation of the
pattern that X-ray shocks associated with hotspots appear to be
upstream of the radio synchrotron shocks.  Studies of other classical
double radio sources have revealed evidence of previous epochs of jet
ejection namely (i) double-double radio galaxies \citep[e.g.\
][]{Schoenmakers2000}, (ii) the double-double X-ray/radio galaxies by
\citep{Laskar2010} and (iii) the presence of a relic X-ray counter-jet
in the prototypical classical double radio galaxy, Cygnus\,A
\citep{Steenbrugge2008}.  In this study, of the giant ($\sim$0.5\,Mpc)
and high-redshift ($z \sim 2.5$) radio galaxy of 4C23.56, two strands
of evidence have emerged that point to there being a previous episode
of jet activity in this object other than the episode delineated by
the current radio synchrotron emission.  These strands of evidence
are: (i) outside of the hotspots, further away from the nucleus, there
is extended radio emission that may be plausibly explained by fast
particles, freshly accelerated at the hotspots, leaking into
pre-existing, weakly magnetised plasma that may be the relic lobes of
a previous epoch of jet activity and (ii) the radio spectrum is
considerably harder than that of the inverse-Compton emission.  This
argues against the emission being drawn from one single population of
relativistic particles.  A simple explanation for this is that the
radio synchrotron is from more recently accelerated particles and the
softer spectrum of the ICCMB emission corresponds to a previous
episode of jet activity, that may even have been re-energized
following compression because of the propagation of the current
episode of jet ejection.

\section*{Acknowledgements}

KMB and ACF thank the Royal Society.  This research has made use of
the NASA/IPAC Extragalactic Database (NED) which is operated by the
Jet Propulsion Laboratory, California Institute of Technology, under
contract with the National Aeronautics and Space Administration.

\bibliographystyle{mn2e} 

\end{document}